# Density functional theory study of the structure and energetics of negatively charged pyrrole oligomers

Yafei Dai[1], Sugata Chowdhury[1], Estela Blaisten-Barojas[1,2]

[1]Computational Materials Science Center, George Mason University, MS 6A2,

Fairfax, Virginia 22030, USA

[2]Department of Computational Sciences, George Mason University, MS 6A2,

Fairfax, Virginia 22030, USA

**Abstract:** First-principles calculations are used to investigate the electronic properties of neutral and negatively charged n-pyrrole oligomers with n= 2-18. Chains of neutral oligomers are bent while the negatively charged oligomers become almost flat due to accumulation of negative charge at the end monomers. Several isomers of the short oligomers (n < 6) display negative electron affinity, although they are energetically stable. For longer oligomers with n ≥ 6, the electron affinity turns positive, increasing with oligomer length. The doping of 12-pyrrole with lithium atoms is studied, showing that negative oxidation states are possible due to charge transfer from dopant to oligomer at locations close to dopant. These molecular regions support extra negative charge and exhibit a local structural change from benzenoid to quinoid in the C-C backbone conjugation. Additional calculations of neutral and doped polypyrrole are conducted showing that the doped infinite polymer chain displays a substantial reduction of the energy band gap and the appearance of dopant-based bands in the gap.

*Email address:* blaisten@gmu.edu .

# 1 Introduction

Since their discovery thirty years ago, research on conducting polymers has been at the frontier [1]. Among conducting polymers, polypyrrole (PPy) is considered to be a prototype conjugated polymer that displays unique electrical, mechanical, and physical properties. These peculiar properties are exploited in a variety of applications from photonic devices to mechanical actuators. The electrical properties of PPy-based devices are tailored for a particular application through the choice of appropriate dopants [2]. PPy is synthesized by electrochemical methods [3] in two different phases: reduced (electrically neutral) and oxidized (electrically charged) when electrons are transferred to the oxidizing agent [4,5]. There are two ways to reach the reduced phase via dopants. One mechanism consists of oxidizing the polymer with electronegative dopants that acquire electrons from the polymer. In this case reduction occurs when electrons are transferred to the polymer from different dopants or from counterions in the solvent. The second mechanism occurs when electropositive dopants donate electrons to the polymer and reduction occurs when the polymer returns the extra negative charge to additional dopants in the polymeric matrix. Therefore, doped PPy may lead to positive or negative oxidation states.

Much less is known experimentally about pyrrole oligomers with n monomers (n-Py), both neutral and oxidized or doped with electropositive or electronegative compounds. Over the years, a series of theoretical studies have provided different predictions for pyrrole oligomers. For example, the structure of bipyrrole has been studied at various levels of approximation [6,7], vibrational spectra have been predicted within Hartree Fock and semi-empirical approaches [8-11] and within density functional theory (DFT) approaches [7], changes in the UV and visible absorption spectra of neutral and positively charged oligomers due to dopants have been put forward [12], the effect of counterions on oligomer structure has been investigated [13], excited states of pyrrole and bipyrrole are published [14], and attempts to calculate ionization potentials and electron affinitites with simplified methodologies have continued to be active [15].

This paper focuses on studying the electronic and structural properties of negatively charged n-Py oligomers, their electron affinity, and structural changes in lithium-doped n-Py. Conjugated oligomers may be doped with electronegative or electropositive atoms or molecules. In the former case, there is electron transfer from the oligomer to the dopant while in the later case electron transfer occurs from the dopant to the oligomer. These charge transfer mechanisms produce structural changes in the oligomer backbone that might be different if the oligomer accepts electrons than when it donates them. In previous studies we addressed the behavior of n-Py oligomers donating negative charge to the dopants [7]. In this paper we analyze structural changes occurring in the n-Py backbone due to lithium atom dopants. Calculations are done at the all-electron DFT approach with gradient correction as in our previous work [7]. Because calculation of electron affinities requires basis sets with polarization and diffuse functions, three triple-z basis sets are considered. This paper is organized as follows. Section 2 describes the computational details for obtaining the electronic structure. Section 3 investigates the structure, energetics and vibrational analysis of neutral and anion bipyrrole, tripyrrole and tetrapyrrole possible isomers. Additionally, the isomerization reactions in 3-Py and 4-Py from *anti-gauche* to *syn-gauche* structures are discussed in this section. Energetics and electron affinity of n-Py neutral and anion oligomers with n ≥ 6 are reported in Section 4.

This section contains a study of lithium doped 12-Py and the band structure of PPy by considering an infinite chain based on periodic boundary conditions of a 4-Py motif. Section 5 concludes this paper.

**2 Methods**

All-electron DFT within the Becke three-parameter hybrid approach [16] including local and non-local correlation functionals as implemented in Gaussian0*3* [17] is adopted throughout this study. The choice of the correlation functional was based on comparative results of pyrrole (Py) stucture using a variety of functionals: B3PW91 [18,19], B3LYP [20–22], B3BMK [23], M06-HF [24]. Results closest to experiment [25] are obtained with B3PW91 reproducing the experimental $C_{2v}$ planar structure of neutral Py monomer. The B3PW91 approach is then adopted which additionally allow for comparison to our previous studies [7]. In the study of doped n-Py oligomers where a considerable charge transfer takes place, comparison between results within B3PW91 and full exchange M06-HF approaches is reported.

Structures of oligomers are optimized using triple valence basis sets 6-311G, 6-311++G [26] with diffuse functions [27] and double valence basis sets 6-31+G (3d, 3p) with diffuse and polarization functions [28] for all atoms. Comparison between the optimal calculated structure of neutral Py and experiment [25] indicates that these three basis sets yield comparable relative errors in the geometry: 0.38% with 6-311G, 0.42% with 6-311++G and 0.37% with 6-31+G (3d,3p). The Py monomer has a strong dipole moment along the direction of the N-H bond (Y-axis) of 1.89 D with the 6-311++G basis sets (1.85 D with 6-31+G (3d,3p)). The quadrupole matrix is diagonal in a set of axis where the X-axis is perpendicular to the N-H bond and contained in the plane of the molecule and the Z-axis is perpendicular to the molecular plane passing through the center of the Py ring. The quadrupole matrix in this set of axis is diagonal with the XX, YY, ZZ matrix elements having values of -27.26, -24.11, -34.97 DÅ using 6-311++G and -27.55, -24.32, -34.33 DÅ using 6-31+G (3d,3p). These properties are in excellent agreement with our previous results employing the 6-311G basis set without diffuse or polarization functions [7].

Geometry optimization for each n-Py oligomer is attained by minimizing the molecular electronic energy with respect to coordinates of all atoms in 3D using the Berny algorithm and redundant internal coordinates [29, 30]. For n-Py optimized geometries, the vibrational analysis is routinely performed to ensure the existence of a minimum. For transition states (TS), the displacement of atoms corresponding to the vibrational mode associated with the imaginary frequency is checked to ensure that the TS geometry is attained. Furthermore, the intrinsic reaction coordinate (IRC) method allows us to assess that the TS connects the two desired minima (two different isomers) along the potential energy surface [31,32]. Solvent effects are included for the small oligomers with the polarized continuum model (PCM) approach [33].

In forthcoming sections, binding energies E are reported with respect to the separated atoms energies. The electron affinity EA is calculated as a difference between all electron total energies as

$$EA = E_{total}(neutral) - E_{total}(anion) \qquad (1)$$

where $E_{total}$(neutral) is the total energy of the optimized neutral oligomer and $E_{total}$(anion) is the total energy of the optimized negatively charged oligomer. Singly negative charged n-Py oligomers are referred to as anions in the following sections.

**3 Bipyrrole, Tripyrrole and Tetrapyrrole Anions**

The geometrical structure of bipyrrole, as well as that of longer oligomers, is modulated the most by the monomer rotational degree of freedom around the C-C bond joining contiguous monomers. Neutral bipyrrole (2-Py) has four stable isomers: two puckered structures *anti-gauche, syn-gauche* and two planar structures *anti, syn* [7]. These four neutral isomers are stable irrespective of the basis sets used. The *anti-gauche* isomer is the lowest in energy with a torsion angle around 150° followed by the *syn-gauche* isomer that lies a few hundredths eV above as reported in Table I and has a torsion angle around 50°. The cation 2-Py$^+$ displays two planar isomers: *anti* ($C_{2h}$) and *syn* ($C_{2v}$) and the transition between them is not thermally possible [7]. Our calculations of the anion 2-Py$^-$ demonstrate that the negatively charged molecule has also two possible planar isomers: *anti* ($C_{2h}$) and *syn* ($C_{2v}$). The lowest energy isomer of 2-Py$^-$ is the $C_{2v}$ structure and the $C_{2h}$ structure is higher in energy as seen from values of the binding energies given in Table I. With the triple-valence basis sets the transition state is 0.20 eV above the ground state of the anion $C_{2v}$ isomer (0.26 eV with 6-31+G (3d, 3p)). The TS corresponds to a high-energy torsion barrier separating the two isomers, eliminating the possibility of thermally induced isomerization. Both anion isomers have electronic energies above their neutral counterparts. Consequently the electron affinity is negative as occur in molecular anion where the extra electron is dipole-bound [34, 35]. Thus, we predict that 2-Py$^-$ will not be observed experimentally when a fairly long lifetime of the anion is required for performing the measurement because of electron recombination with the surroundings.

The IR-active normal mode frequencies of 2-Py are distributed in two frequency domains at 600 to 1600 $cm^{-1}$ and at 3200 to 3700 $cm^{-1}$ for the four neutral and anion isomers. This is also the case for the cation isomers [7]. For comparison purposes the vibrational spectra obtained with 6-311++G and 6-31+G (3p,3d) are discussed here. Most vibrational frequencies in the 2-Py$^-$ isomers are comparable to those in the neutral isomers displaying small red shifts with the exception of two modes that become substantially more intense in the anion isomers. Additionally, vibrational mode intensities in each spectrum are referred to the peak with maximum intensity of that spectrum. The N-H stretching mode in the *anti-gauche* neutral isomer located at 3691 $cm^{-1}$ with intensity of 0.4 (3675 $cm^{-1}$, 0.6 intensity with 6-31+G (3d, 3p)) is red-shifted to 3557 $cm^{-1}$ with 1.0 intensity (3430 $cm^{-1}$, 1.0 intensity with 6-31+G (3d, 3p)) in 2-Py$^-$ $C_{2h}$ anion. The same mode in the *syn-gauche* neutral isomer, located at 3685 $cm^{-1}$ with 0.12 intensity (3669 $cm^{-1}$ 0.16 intensity with 6-31+G (3d, 3p)) is red-shifted to 3410 $cm^{-1}$ with 0.82 intensity (3292 $cm^{-1}$, 0.97 intensity with 6-31+G (3d, 3p)) in the 2-Py$^-$ $C_{2v}$ anion. Similarly, the hydrogens rocking mode in the N-H and C-H bonds in the *anti-gauche* neutral isomer located at 1134 $cm^{-1}$ with 0.32 intensity (1126 $cm^{-1}$, 0.32 intensity with 6-31+G (3d, 3p)) is red-shifted to 1122 $cm^{-1}$ with 0.42 intensity (1099 $cm^{-1}$, 0.29 intensity with 6-31+G (3d, 3p)) in the 2-Py$^-$ $C_{2h}$ anion. The same mode in the *syn-gauche* neutral isomer, located at 1143 $cm^{-1}$, has an intensity of 0.16 (1131 $cm^{-1}$, 0.16 intensity with 6-31+G (3d, 3p)) while the $C_{2v}$ 2-Py$^-$ is red-shifted to 1134 $cm^{-1}$ with 1.0 intensity (1102 $cm^{-1}$, 1.0 intensity with 6-31+G (3d, 3p)).

Combinations of *anti-gauche* with *syn-gauche* orientations of monomers in tripyrrole (3-Py⁻) and tetrapyrrole (4-Py⁻) give rise to three and six isomers, respectively. The three neutral isomers of 3-Py are shown in Figure 1a. The ground states of these neutral isomers are singlet electronic states. Binding energies are reported in Table I, including comparison of results calculated with the three basis sets considered in this study. The vibrational analysis of isomers and their anion counterpart yields positive frequencies, indicating that the values reported in Table I correspond to minima of the potential energy surface. The 3-Py isomer of lowest energy, isomer I, has two 150° torsion angles (157° with 6-31+G (3d, 3p)) between monomer planes containing the N-H bond. Isomer III is the highest energy 3-Py isomer that has two 43° torsion angles between monomer planes (39° with 6-31+G (3d, 3p)). The two torsion angles in isomer II are 150° and 43° (157°, 39° with 6-31+G (3d, 3p)). Two transition states, $TS_1$ and $TS_2$, lie between these three isomers ground states at 0.11 eV and 0.17 eV (0.13, 0.15 eV using 6-31+G (3d, 3p)). The first transition structure has the third ring in isomer I rotated 90° while the second transition structure has the third ring of isomer II rotated 90°. Schematics of the potential energy along the isomerization path are depicted in Figure 1b. Solvent effects, assuming a relative dielectric constant for water of 78.355, were considered. The overall effect is to stabilize the isomers and destabilize the transition states as indicated in parenthesis in Figure 1. Geometry optimization of 3-Py⁻ isomers yields almost planar structures in doublet states. Table I contains the binding energies of the three anion isomers corresponding to geometry-optimized structures. Isomer III, with the three N-H bonds pointing in the same side of the chain, is the most stable 3-Py⁻. This isomer has a marginal (almost zero) electron affinity indicating that under favorable circumstances, a weak electron attachment to 3-Py may be observed due to a dipole binding mechanism [34,35]. However, the most stable neutral isomer I would need to overcome a 0.17 eV (0.15 eV with 6-31+G (3d, 3p)) isomerization barrier to transform into isomer III before an electron could be attached. Therefore this process might be feasible at high temperatures only.

For tetrapyrrole there are six isomers combining the *anti* and *syn* orientations of the N-H bonds. For simplicity these isomers are represented with up/down arrows. States and binding energies are reported in Table I, which correspond to a full geometry optimization of each isomer and each anion. The vibrational analysis of all isomers and their anions yield positive frequencies indicating that the values reported correspond to minima of the potential energy surface. The 4-Py neutral isomer ↑ ↓ ↑ ↓ (torsion angles 150°, 151°, 150° with 6-311++G and 156°, 157°, 156° with 6-31+G (3d, 3p)) is the most stable and the isomer ↑ ↑ ↑ ↑ (41°, 33°, 41° with 6-311++G and 35°, 30°, 35° with 6-31+G (3d, 3p)) is the least stable. The other four neutral isomers are ordered in increasing energy order as ↑ ↓ ↓ ↑ (149°, 31°, 149° with 6-311++G and 155°, 27°, 155° with 6-31+G (3d, 3p)), ↑ ↓ ↑ ↑ (150°, 153°, 40° with 6-311++G and 156°, 158°, 36° with 6-31+G (3d, 3p)), ↑ ↑ ↓ ↓ (41°, 152°, 41° with 6-311++G and 36°, 157°, 36° with 6-31+G (3d, 3p)), ↑ ↑ ↑ ↓ (40°, 35°, 149° with 6-311++G and 36°, 31°, 156° with 6-31+G (3d, 3p)).

The most stable 4-Py⁻ anion is the ↑ ↑ ↑ ↑ isomer. This is an almost planar molecule with torsion angles 4°, 6°, 4° (3°, 4°, 3° with 6-31+G (3d, 3p)). This isomer displays a marginal almost zero electron affinity presumably identifying a dipole-bound anionic state. All other anion isomers are less stable than their corresponding neutral isomer as shown in Table I. Two transition states of neutral 4-Py were found to lie at 0.11 eV (0.14 with 6-31+G (3d, 3p)) above the ground state energy of the ↑↓↑↓ isomer to reach the ↑↓↑↑ isomer and 0.24 eV above the ground state (0.29 eV with 6-31+G(3d, 3p)) to reach the ↑↑↑↑ isomer. We predict that in order to detect experimentally 4-Py$^-$, this oligomer needs first to isomerize, overcoming a barrier of about a quarter eV, and only then an electron would be attachable to the ↑↑↑↑ isomer. This process may occur at very high temperatures.

Both neutral and anion isomers of tripyrrole and tetrapyrrole isomers display IR-active vibrational modes in the same frequency regions as bipyrrole. It is interesting to note that the anion ↑↑↑↑ isomer has three intense IR-active lines within 3330 - 3420 cm$^{-1}$ with relative intensities 1.0, 0.4, 0.7 that originate from the stretching of the N-H bonds in monomers in-between the oligomer ends. These modes are red shifted from the 3691 cm$^{-1}$ mode in the neutral isomer ↑↓↑↓ that has an intensity of 0.3. Additionally, the neutral isomer ↑↑↑↑ displays a fairly intense mode around 1600 cm$^{-1}$ corresponding to hydrogen-rocking motion that is absent in the corresponding anion.

**4 Larger n-Py anions (n = 6-9, 12, 15, 18) and PPy with dopants.**

Full geometry optimization of n-Py is performed for sizes with n = 6, 8, 9 for both neutral and anions with the three basis sets considered. Only neutral and anion isomers with alternating direction of the N-H bond are analyzed for these larger oligomers due to the high computational cost. For oligomers with n = 12, 15, 18 the geometries of neutral and anion oligomers are fully optimized with the 6-311G basis sets and only single point calculations with the 6-311++G and the 6-31+G (3d,3p) are reported. The reason for this is that comparison of the smaller oligomers optimized structures with the three basis sets shows very small differences in bond lengths and angles calculated, with a consistent decrease in difference as the oligomer length increases. Table II summarizes the binding energies per monomer calculated with this scheme and the three basis sets. Electron affinities of these oligomers are reported in Table II, clearly showing a steady increase with increasing oligomer length. Based on these data, EA increases steadily as a power law: $0.187(n-6)^{1/2}$.

The polymeric matrix of PPy is usually grown in an ionic solution where the growing polymer chains may gain electrons from electropositive dopants. Therefore, the charge distribution along the conjugated polymer chain is an important property of the anion oligomers. Based on the Muliken population analysis, the charge on each C atom in the conjugated chain for 12-Py with Li atom dopants is investigated in this work. In the oligomer anion the negative charge tends to be pushed to the two end monomers and in doing so the oligomer becomes more planar as more negative charge is transferred to the molecule from the dopants. Similar to the case of positively charged 12-Py [7], for multiply -charged anionic oligomers there is a structural change in the conjugated backbone chain from the benzenoid structure (single C-C bond between contiguous monomers) to the quinoid structure (double C-C inter-monomer bond) in chain regions neighboring dopants. Therefore, when electropositve dopants such as Li atoms are in the proximity of n-Py oligomers, a transfer of negative charge to the polymer occurs in their immediate proximity. The effect of Li dopants on the alternation of single-double C-C bonds along the conjugated chain of 12-Py is illustrated in Figure 2, where $\Delta l_{C-C}$ is the change in length of C-C bonds and $\Delta q_C$ is the change in charge on the C atoms occurring in the doped oligomer with respect to the oligomer without dopants. Calculation for this case is based on a full optimization of the geometry of 12-Py with three Li dopants using the 6-311G basis sets. Reported values of energies using the other basis sets correspond to one-point calculation considering that optimized geometry of the doped system. Plots on the left of the figure correspond to the B3PW91/6-31+G (3d, 3p) results, which we compare with our calculation obtained under the HF-M06 exchange DFT approach depicted on the right. These results correspond to a full optimization of the doped system geometry. It is clear from the figure that, irrespective of the method, changes occur preferentially at chain regions where the polymer sustains a negative charge due to electron transfer from dopants in that vicinity. The Li dopants are about 1.9 Å away from the chain, located above and below two contiguous monomers of a 4--monomer motif, and the 4-monomer motif repeats three times in 12-Py as depicted in the top of Figure 2. This configuration is proved to correspond to a maximum charge transfer of 1.36 e per motif (1.28 e/motif with M06-HF) by inspecting several energy minima of the potential energy surface.

Neutral PPy is an insulator and when becoming positively charged one localized state appears in the gap per removed electron [7]. As part of the present work, additional calculations within the B3PW91/6-311G approach are performed for the infinite polymer PPy, both pure and doped with Li atoms. Calculations are done with periodic boundary conditions in the X-direction considering a repeatable 4-monomer motif in the planar configuration. With this motif the unit cell of length is 14.35Å. In the case of doped PPy, the unit cell contains the same 4-monomer motif as in pure PPy plus two Li atoms located one above and one below (at 1.9 Å) two contiguous monomers. Figure 3 illustrates a comparison between the one-electron level spectrum of 12-Py and the band structure of PPy. Top plots depict systems without dopants while the bottom plots show the effect of Li-doping. Energies are referred to the HOMO in the case of 12-Py and to the Fermi energy in PPy. The Fermi energy $E_F$ value solves the equation that equates the number of electrons N to the sum of electron occupation probabilities (Fermi functions) of eigenstates composing the bands:

$$N = \sum_{\alpha=1}^{n_{band}} \sum_{k=1}^{n_k} \frac{2}{1+e^{\frac{(E_{\alpha,k}-E_F)}{K_B T}}} \qquad (2)$$

where $n_{band}$ = 16 is the number of bands, $n_k$ = 81 is the number of k-points in each band allowing up to 162 electrons per band, $K_B$ is the Boltzman constant, T = 600 K is a broadening temperature and N = 2 $n_k$ $n_{band}$. Values of these parameters ensure accuracy of 0.01 eV in the determination of $E_F$.

In the discrete spectrum of 12-Py, Li dopants give rise to a manifold of states in the proximity of the HOMO located at -3.50 eV. These states are tightly bundled placed within the 3.36 eV HOMO-LUMO energy gap of neutral 12-Py. Correspondingly, the band structure of PPy has $E_F$ = -3.52 eV and a band gap of 3.2 eV. As illustrated in Figure 3, Li-doped PPy has $E_F$ = -1.34 eV, a strongly depleted band gap of 2.4 eV and two new narrow bands with widths 0.001 eV and 0.542 eV located in the gap. A charge transfer of 1.2 e takes place from the Li-dopants toward the 4-monomer motif in the PPy chain. At T = 600 K the total electron occupation of the narrow band below $E_F$ is about 158 with four holes, while the band just above $E_F$ has an electron occupation of about 4.

At T = 300 K the band below $E_F$ is basically filled (161.7). We predict that more electropositive dopants will lead to a conducting system in which bands in the neighborhood of $E_F$ are closer to each other.

**5 Conclusion**

This paper reports a thorough study of neutral and anion n-Py oligomers (n=2-18) at the density functional theory level with large basis sets. The study confirms earlier structure and energetics results within DFT and smaller basis sets for neutral oligomers [7] and puts forward the new structure and energetics findings for their anions. Short n-Py oligomers with n <6 attach an electron through dipole-binding. The resulting anions are less stable than their neutral counterparts, and thus may be difficult to detect experimentally. However, larger oligomers have positive EAs that increase with polymer length. It is demonstrated that in oligomers doped with electropositive dopants the electron transfer mechanism occurs around n-Py monomers in the neighborhood of the dopants. Lithium atom dopants create an energy-localized bundle of states in the HOMO-LUMO gap of the neutral oligomer. The corresponding band structure of PPy with Li dopants shows the appearance of dopant bands in the energy gap of neutral PPy accompanied by a substantial decrease in the band gap.

**Acknowledgments**

We acknowledge the National Science Foundation grant CHE0626111 for partial support and the TERAGRID grant PHY050026T for the computer time allocation.

**References**

1. Chiang, C. K.; Fincher, C. R.; Park, Y. W.; Heeger, A. J.; Shirakawa, H.; Louis, E. J.; Gau, S. C.; MacDiarmid, A. G. Phys Rev Lett 1977, 39, 1098.

2. Vermeir, I. E.; Kim, N. Y.; Laibinis, P. E. Appl Phys Lett 1999, 74, 3860.

3. Yamaura, M.; Hagiwara, T.; Iwata, K. Synth Met 1988, 26, 209.

4. Smela, E. J Micromech Microeng 1999, 9, 1.

5. Smela, E.; Gadegaard, N. Adv Mater (Weinheim, Ger.) 1999, 11, 953.

6. Bredas, J. L.; Street, G. B.; Themans, B.; Andre, J. M. J Chem Phys 1985, 83, 1323.

7. Dai, Y.; Blaisten-Barojas, E. J Chem Phys 2008, 129, 164903.

8. Tian, B.; Zerbi, G. J Chem Phys 1990, 92, 3886.

9. Kofranek, M.; Kovar, T.; Karpfen, A.; Lischka, H. J Chem Phys 1992, 96, 4464.

10. Rabias, I.; Howlin, B. J. Comput Theor Polym Sci 2001, 11, 241.

11. Lee, S. Y.; Boo, B. H. J Phys Chem 1996, 100, 15073.

12  Okur, S.; Salzer, U. J Phys Chem A 2008, 112, 11842.

13 Tamm, T.; Tamm, J.; Karelson, M. Int J Quantum Chem 2002, 88, 296.

14. Burcl, R.; Amos, R. D.; Handy, N. C. Chem Phys Lett 2002, 355, 8.

15 Colle, R.; Montagnani, R.; Salvetti, O. Theor Chem Acc 1999, 101, 262.

16. Becke, A. D. J Chem Phys 1993, 98, 5648.

17. Frisch, M. J.; Trucks, G. W.; Schlegel H. B. et. al. Gaussian 03, Revision C.01, and Gaussian 09, Revision A.02, Gaussian, Inc., Wallingford, CT, 2004.

18. Perdew, J. P.; Chevary, J. A.; Vosko, S. H.; Jackson, K. A.; Pederson, M. R.; Singh, D. J.; Fiolhais, C. Phys Rev B 1992, 46, 6671.

19. Perdew, J. P.; Burke, K.; Wang, Y. Phys Rev B 1996, 54, 16533.

20. Stephens, P.J.; Devlin, F. J.; Chabalowski,, C. F.; Frisch, M. J. J Phys Chem 1994, 98, 11623.

21. Becke, A. D. Phys Rev A 1988, 38, 3098.

22. Lee, C.; Yang, W.; Parr, R. G. Phys Rev B 1988, 37, 785.

23. Boese, A. D.; Martin, J. M. L. J Chem Phys 2004, 121, 3405.

24. Zhao, Y.; Truhlar, D. G. J Phys Chem A 2006, 110, 13126.

25. Nygaard, L.; Nielsen, J. T.; Kirchheiner, J.; Maltesen, G.; Rastrup-Andersen, J.; Sorensen, G. O. J Mol Struct 1969, 3, 491.

26. Krishnan, R.; Binkley, J. S.; Seeger, R.; Pople, J. A. J Chem Phys 1980, 72, 650.

27. Clark, T.; Chandrasekar, J.; Spitznagel, G. W.; Schleyer, P. V. R. J Comp Chem 1983, 4, 294.

28. Frisch, M. J.; Pople, J. A.; Binkey, J. S. J Chem Phys 1984, 80, 3265.

29. Reed, A. E.; Weinhold, F. J Chem Phys 1983, 78, 4066.

30. Peng, C.; Ayala, P. Y.; Schlegel, H. B.; Frisch, M. J. J Comp Chem 1996, 17, 49.

31. Fukui, K. J Phys Chem 1970, 74, 4161.

32. Gonzalez, C.; Schlegel, H. B. J Chem Phys 1989, 90, 2154.

33. Cances, M.; Mennucci, B.; Tomasi, J. J Chem Phys 1997, 107, 3032.

34. Gutowski, M.; Jordan, K.; Skurski, P. J Phys Chem A 1998, 102, 2624.

35 Gutowski, M.; Skurski, P.; Boldyrey, A. I.; Simons, J.; Jordan, K. D. Phys Rev A 1996, 54, 1906.

**Figure Captions**

Figure 1. Potential energy surface of tripyrrole isomers along the isomerization path calculated within B3PW91/6-311++G.

Figure 2. Changes of C-C bond length and charge on C atoms along the conjugated chain in 12-Py$^-$ anion with respect to neutral 12-Py. Large circles identify inter monomer locations.

Figure 3. Energy spectrum of 12-Py compared to the band structure of PPy for systems without dopants (top) and with Li dopants (bottom). The 12-Py spectrum is obtained with 6-31+G (3d,3p) basis sets and the PPy bands with 6-311G basis sets.

Table I. Binding energies of n-Py and n-Py⁻ stable isomers with n = 2, 3, 4. Energies are referred to the separated atoms energies: -13.7133 eV for H, -1027.4931 eV for C, -1481.9450 eV for N with the 6-311G; -13.7156 eV for H, -1027.5601 eV for C, -1482.0417 eV for N with the 6-311++G; -13.6642 eV for H, -1027.4197 eV for C, -1482.0417 eV for N with the 6-31+G (3d,3p) basis sets.

| | | | neutral | | | anion | | | |
|---|---|---|---|---|---|---|---|---|---|
| | | | 6-311G | 6-311++G | 6-31+G (3d, 3p) | | 6-311G | 6-311++G | 6-31+G (3d,3p) |
| n | isomer | state | E/n (eV) | | | state | E/n (eV) | | |
| 2 | ↑↓ | $C_2, {}^1A$ | -53.4209 | -53.1300 | -54.9775 | $C_{2h}, {}^2A_g$ | -52.6860 | -52.9188 | -54.6165 |
| | ↑↑ | $C_2, {}^1A$ | -53.3599 | -53.0954 | -54.9399 | $C_{2v}, {}^2A_1$ | -52.6994 | -52.9600 | -54.6319 |
| 3 | ↑↓↑ | $C_1, {}^1A$ | -52.6129 | -52.3189 | -54.1434 | $C_1, {}^2A$ | -52.3243 | -52.1983 | -53.9450 |
| | ↑↑↓ | $C_1, {}^1A$ | -52.5636 | -52.2946 | -54.1304 | $C_1, {}^2A$ | -52.3362 | -52.2122 | -53.9570 |
| | ↑↑↑ | $C_1, {}^1A$ | -52.5673 | -52.2735 | -54.1167 | $C_1, {}^2A$ | -52.3448 | -52.2743 | -54.0018 |
| 4 | ↑↓↑↓ | $C_2, {}^1A$ | -52.2097 | -51.9145 | -53.7274 | $C_1, {}^2A$ | -51.0819 | -51.8370 | -53.6606 |
| | ↑↓↓↑ | $C_2, {}^1A$ | -52.1968 | -51.9007 | -53.7040 | $C_1, {}^2A$ | -51.0880 | -51.8411 | -53.6541 |
| | ↑↓↑↑ | $C_1, {}^1A$ | -52.1922 | -51.8974 | -53.6994 | $C_1, {}^2A$ | -51.0825 | -51.8358 | -53.6476 |
| | ↑↑↓↓ | $C_2, {}^1A$ | -52.1793 | -51.8844 | -53.6851 | $C_1, {}^2A$ | -51.0821 | -51.8342 | -53.6464 |
| | ↑↑↑↓ | $C_1, {}^1A$ | -52.1784 | -51.8827 | -53.6837 | $C_1, {}^2A$ | -51.0912 | -51.8423 | -53.6534 |
| | ↑↑↑↑ | $C_2, {}^1A$ | -52.1582 | -51.8630 | -53.6733 | $C_1, {}^2A$ | -51.0916 | -51.9241 | -53.6779 |

Table II. Binding energy and electron affinity of neutral and anion n-Py oligomers with n = 6, 8, 9, 12, 15, 18. Binding energies are referred to energies of the isolated atoms (given in Table I caption).

| | | neutral | | | | anion | | | | | |
| | | 6-311G | 6-311++G | 6-31+G(3d,3p) | | 6-311G | | 6-311++G | | 6-31+G (3d,3p) | |
| n | state | | E/n (eV) | | state | E/n (eV) | EA(eV) | E/n(eV) | EA(eV) | E/n(eV) | EA(eV) |
|---|---|---|---|---|---|---|---|---|---|---|---|
| 6 | $C_2$ $^1A$ | -51.8064 | -51.5092 | -53.3017 | $C_1$ $^2A$ | -51.7872 | -0.0192 | -51.5164 | 0.0432 | -53.3110 | 0.0557 |
| 8 | $C_2$ $^1A$ | -51.6045 | -51.3070 | -53.0936 | $C_1$ $^2A$ | -51.6175 | 0.0130 | -51.3370 | 0.2396 | -53.1246 | 0.2482 |
| 9 | $C_1$ $^1A$ | -51.5375 | -51.2397 | -53.0244 | $C_1$ $^2A$ | -51.5576 | 0.0201 | -51.2740 | 0.3087 | -53.0595 | 0.3155 |
| 12 | $C_2$ $^1A$ | -51.4029 | -51.1045 | -52.8862 | $C_1$ $^2A$ | -51.4318 | 0.0289 | -51.1431 | 0.4634 | -52.9249 | 0.4642 |
| 15 | $C_1$ $^1A$ | -51.3221 | -51.0234 | -52.8030 | $C_1$ $^2A$ | -51.3528 | 0.0307 | -51.0615 | 0.5709 | -52.8411 | 0.5705 |
| 18 | $C_2$ $^1A$ | -51.2684 | -50.9694 | -52.7474 | $C_1$ $^2A$ | -51.2995 | 0.0311 | -51.0054 | 0.6475 | -52.7835 | 0.6500 |

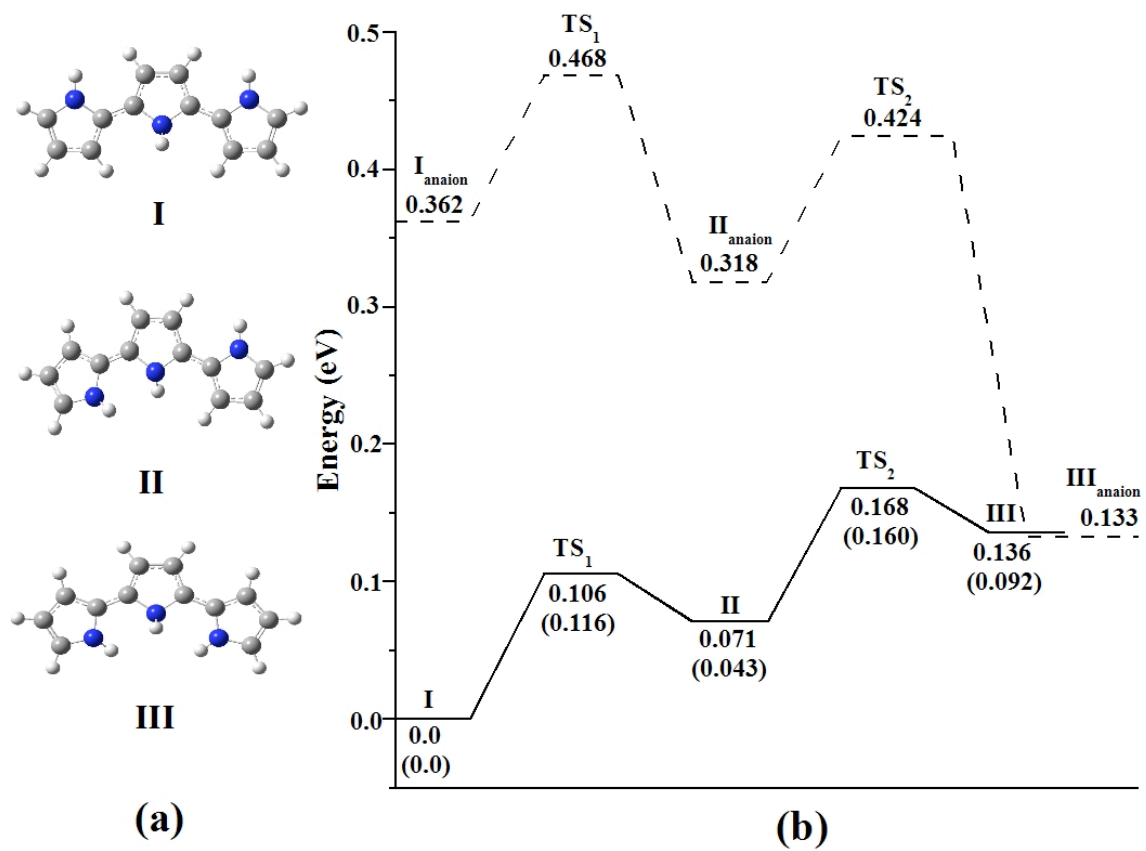

**(a)** **(b)**

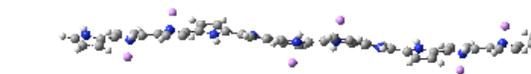
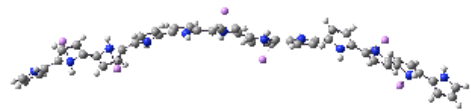
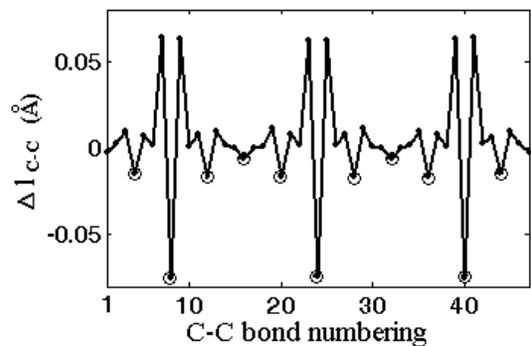
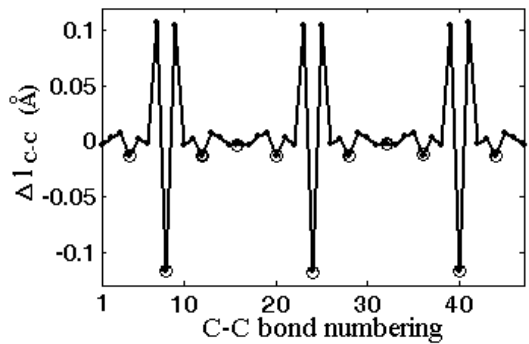
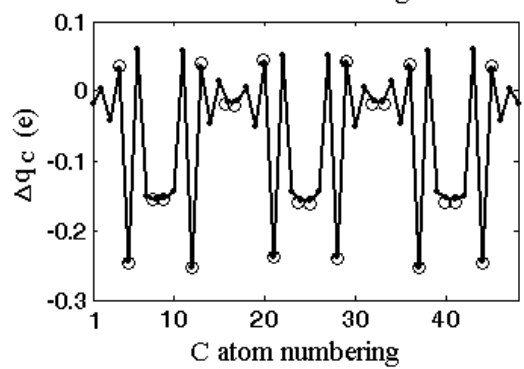
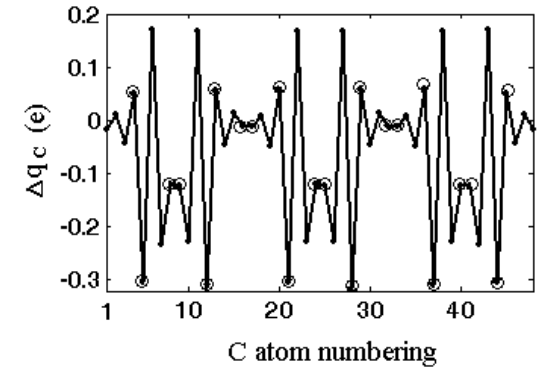

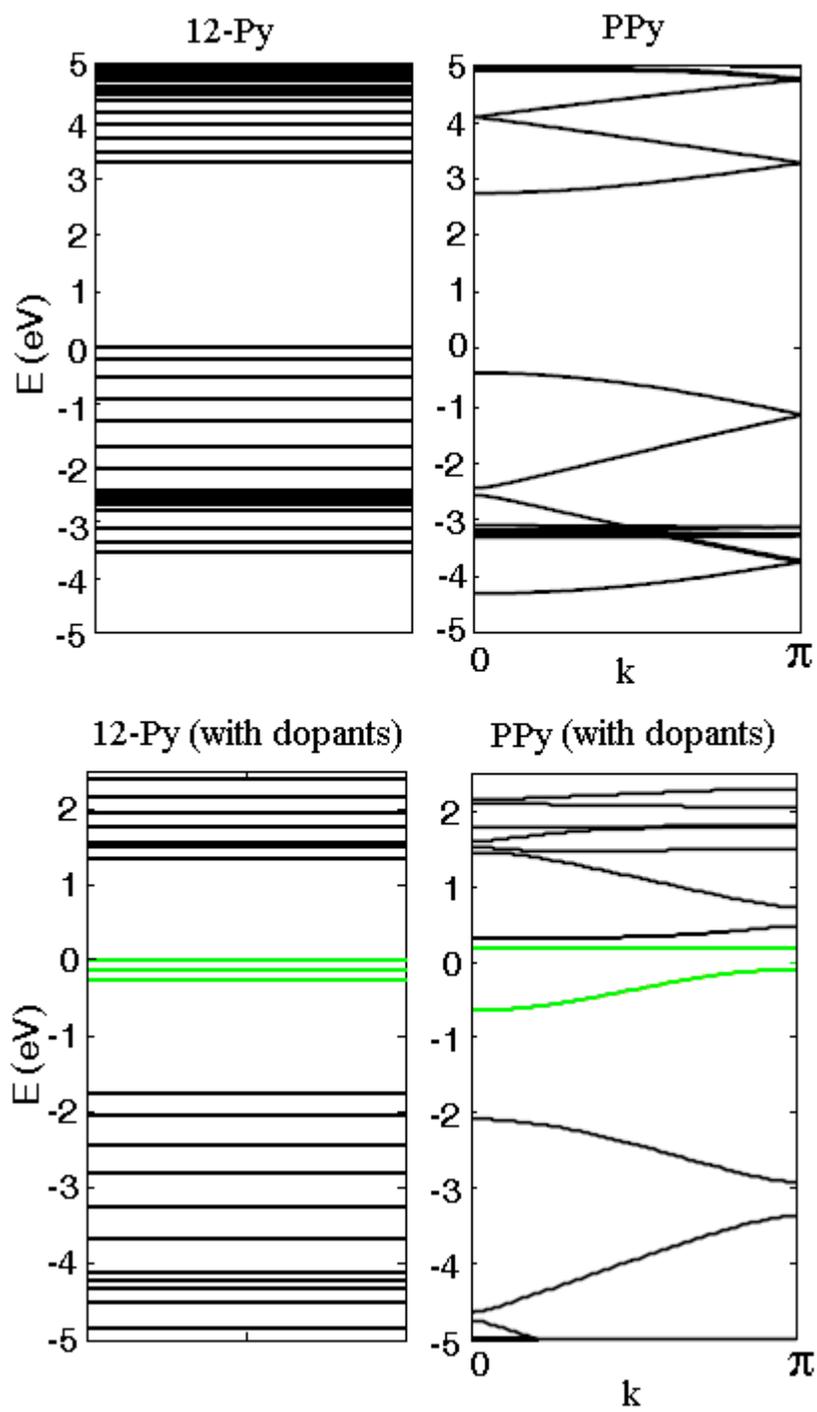
22